\documentclass[twocolumn,prl,floatfix,superscriptaddress,longbibliography]{revtex4-2}

\usepackage{amsmath}
\usepackage{amssymb}
\usepackage[utf8]{inputenc}
\usepackage[T1]{fontenc}
\usepackage{todonotes}
\usepackage{graphicx}
\usepackage{dcolumn}
\usepackage{bm}
\usepackage{xcolor}
\usepackage{lipsum} 
\usepackage{footmisc}
\usepackage{placeins}

\usepackage[colorlinks=true, linkcolor=blue, citecolor=blue, urlcolor=blue]{hyperref}
\usepackage{comment}

\begin{document}

\title{Electrostatic and magnetic control of the anomalous Josephson phase in\\ ferromagnetic hybrid junctions}

\author{D. Rana}
\affiliation{Niels Bohr Institute, University of Copenhagen, 2100 Copenhagen, Denmark}
\author{O. S. Kurapati}
\affiliation{Niels Bohr Institute, University of Copenhagen, 2100 Copenhagen, Denmark}
\author{N. Vanjarapu}
\affiliation{Niels Bohr Institute, University of Copenhagen, 2100 Copenhagen, Denmark}
\author{J. Zhao}
\affiliation{Niels Bohr Institute, University of Copenhagen, 2100 Copenhagen, Denmark}
\author{J. B. F. Wolters}
\affiliation{Niels Bohr Institute, University of Copenhagen, 2100 Copenhagen, Denmark}
\author{Y. Liu}
\affiliation{Niels Bohr Institute, University of Copenhagen, 2100 Copenhagen, Denmark}
\author{P.~Krogstrup}
\affiliation{Niels Bohr Institute, University of Copenhagen, 2100 Copenhagen, Denmark}
\author{A. L. R. Manesco}
\affiliation{Niels Bohr Institute, University of Copenhagen, 2100 Copenhagen, Denmark}
\author{S. D. Escribano}
\affiliation{Department of Condensed Matter Physics, Weizmann Institute of Science, Rehovot 7610001, Israel}
\author{E. Prada}
\affiliation{Instituto de Ciencia de Materiales de Madrid (ICMM), CSIC, 28049 Madrid, Spain}
\author{S. Vaitiek\.{e}nas}
\affiliation{Niels Bohr Institute, University of Copenhagen, 2100 Copenhagen, Denmark}

\date{\today}

\begin{abstract}
    We report phase-sensitive measurements of Josephson junctions formed in InAs nanowires with epitaxial ferromagnetic EuS and superconducting Al shells.
    Using a superconducting interferometer, we observe a continuous, hysteretic anomalous phase shift under axial magnetic-field sweeps.
    A remanent phase bias persists at zero applied field and depends on magnetic history.
    Electrostatic gating strongly modifies the magnitude of the phase response, revealing its sensitivity to the transverse electronic structure of the nanowire.
    Self-consistent electrostatic simulations show that the effective Rashba spin--orbit coupling follows the measured gate-dependent phase response, suggesting its central role in the electrostatic control of the anomalous Josephson phase.
\end{abstract}

\maketitle

The supercurrent through a Josephson junction is a periodic function of the phase difference, $\varphi$, across it~\cite{Josephson1962, Golubov2004}.
Conventional junctions typically favor $\varphi=0$; however, exchange interactions in the weak link can stabilize a $\pi$-shifted ground state~\cite{Buzdin2005, Bergeret2005, Linder2015}.
Transitions between the $0$ and $\pi$ phases have been inferred in metallic junctions from supercurrent dependence on temperature and ferromagnetic-layer thickness~\cite{Ryazanov2001, Kontos2002, Sellier2004, Oboznov2006}, and directly verified by phase-sensitive measurements~\cite{Guichard2003, Bauer2004}.
When magnetic exchange is combined with spin--orbit coupling and broken spatial symmetries, the current--phase relation can instead acquire an anomalous phase shift, $\varphi_0\neq 0,\pi$~\cite{Buzdin2008, Reynoso2008, Yokoyama2014Anomalous, BergeretTokatly2015, Rasmussen2016}.
Such behavior has been observed in semiconductor quantum dots, topological insulators, and semiconductor--superconductor heterostructures~\cite{Szombati2016, Assouline2019, Mayer2020, Strambini2020, Reinhardt2024Link}.

Hybrid InAs nanowires with epitaxial Al shells combine hard induced superconducting gaps with electrostatically tunable Josephson transport~\cite{Krogstrup2015, Chang2015, Goffman2017}. 
Beyond tuning the higher-harmonic content of the current--phase relation~\cite{Spanton2017, Hart2019}, gating can redistribute electronic states across the nanowire cross section~\cite{Antipov2018, Mikkelsen2018, Reeg2018, Escribano2019}.
This redistribution can modify the effective Rashba spin--orbit coupling~\cite{Liang2012, Scherubl2016, Escribano2020} and semiconductor--superconductor hybridization~\cite{Vaitiekenas2018EffectiveG}.
Introducing epitaxial EuS provides ferromagnetic exchange~\cite{Liu2020}, with magnetic proximity strongly dependent on the shell geometry and electronic-state distribution~\cite{Liu2021, Escribano2021, Escribano2022}.
Experiments have demonstrated spin-split superconducting states at zero applied magnetic field~\cite{Vaitiekenas2021, Vaitiekenas2022}, magnetic-domain-dependent superconductivity, including $0$--$\pi$ switching~\cite{Razmadze2023, Zhao2026SpinSplit, Nandi2026}, and field-free Josephson diode behavior~\cite{Telkamp2025FieldFreeDiode}.

Here, we go beyond discrete $0$--$\pi$ switching by demonstrating a continuously evolving anomalous Josephson phase in spin--orbit-coupled ferromagnetic hybrid nanowire junctions.
Using a superconducting interferometer with a nonmagnetic reference junction, we observe a hysteretic evolution of the phase shift under axial magnetic-field sweeps, including a remanent phase bias whose magnitude can be programmed by magnetic history.
We further observe a pronounced gate dependence of the anomalous phase shift, revealing its sensitivity to the electronic structure of the hybrid nanowire.
Self-consistent electrostatic simulations reveal that the effective Rashba spin--orbit coupling follows the same gate-dependent trends as the measured anomalous phase shift, indicating its central role in the electrostatic response.
Together, these results establish ferromagnetic hybrid nanowires as a platform for magnetically programmable and electrostatically tunable Josephson phase biases.

\begin{figure}[t]
    \includegraphics[width=0.48\textwidth]{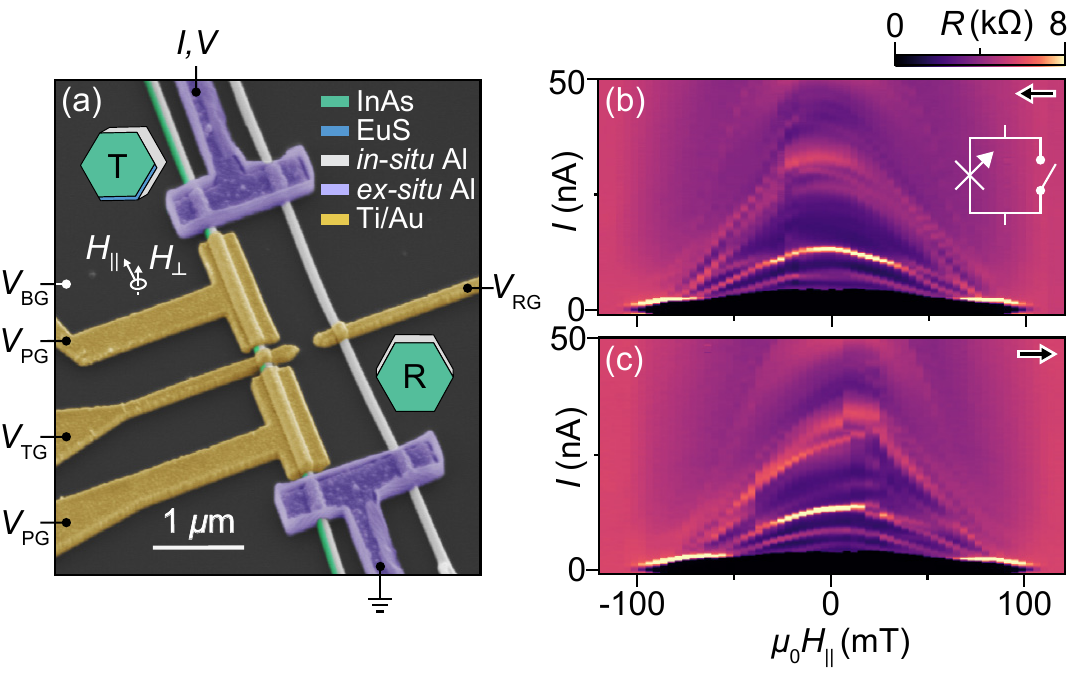}
    \caption{
    (a) Color-enhanced scanning electron micrograph of a superconducting interferometer comprising a ferromagnetic InAs/EuS/Al target junction (T) and a nonmagnetic InAs/Al reference junction (R).
    Gate electrodes enable independent control of junction transparencies and charge-carrier distribution in the target wire.
    (b) and (c) Differential resistance, $R=dV/dI$, of the target junction as a function of current bias, $I$, and axial magnetic field, $\mu_0 H_\parallel$, taken with the reference junction depleted.
    The field is swept from (b) positive to negative and (c) negative to positive, following polarization at $\mu_0 H_\parallel=\pm 200$~mT.
    Data were taken at $V_{\rm BG}=10$~V and $V_{\rm PG}=-5$~V.
    }
    \label{fig:1} 
\end{figure}

The measured superconducting interferometers comprised two nanowires, denoted target (T) and reference (R) [Fig.~\ref{fig:1}(a)].
The target wire consists of a hexagonal InAs core, with epitaxial EuS (2~nm, 2 facets) and Al (20~nm, 3 facets) shells, overlapping on one facet.
The reference wire has a similar structure, but with a thinner Al shell (7~nm) and no EuS.
The nanowires were placed adjacent to each other on a doped Si/SiO$_{\rm x}$ substrate and connected at both ends by \textit{ex situ} Al contacts to form the interferometer loop.
Josephson junctions were defined by selectively removing approximately 100~nm of the Al shells, leaving the EuS intact in the target junction.
The two junctions were independently controlled by local Ti/Au gates deposited on a thin HfO$_{\rm x}$ dielectric layer.
Additional plunger gates tune the carrier density in the target-wire leads from the sides and above, whereas a global back gate tunes the nanowires from below, including the junction regions.
We focus on the measurements from a representative device in the main text, with supporting results from a second device provided in the Supplemental Material~\cite{SupplementalMaterial}.
Four-terminal resistance measurements were performed using standard ac lock-in techniques in a dilution refrigerator equipped with a three-axis vector magnet, at a base temperature of approximately 20~mK.

We first characterize the magnetic-field response of the target junction, with the reference junction depleted.
Before each measurement, the target nanowire was polarized by applying an axial magnetic field $\mu_0 H_\parallel=\pm 200$~mT.
The differential resistance, $R=dV/dI$, exhibits a superconducting window extending approximately between $\mu_0 H_\parallel=\pm 110$~mT for both field-sweep directions [Figs.~\ref{fig:1}(b,c)].
Within this window, hysteretic finite-resistance features appear above the switching current, $I_{\rm SW}$, exhibiting abrupt changes near $\mu_0 H_\parallel=\pm 25$~mT.
The corresponding voltage positions are consistent with multiple Andreev reflections, yielding a superconducting gap $\Delta\approx 110~\mu$eV near zero field, roughly half the bulk Al gap~\cite{SupplementalMaterial}.
The reduced superconducting gap and hysteretic subgap structure highlight the influence of magnetic proximity on superconducting transport in the hybrid junction.

\begin{figure}[t]
    \includegraphics[width=0.48\textwidth]{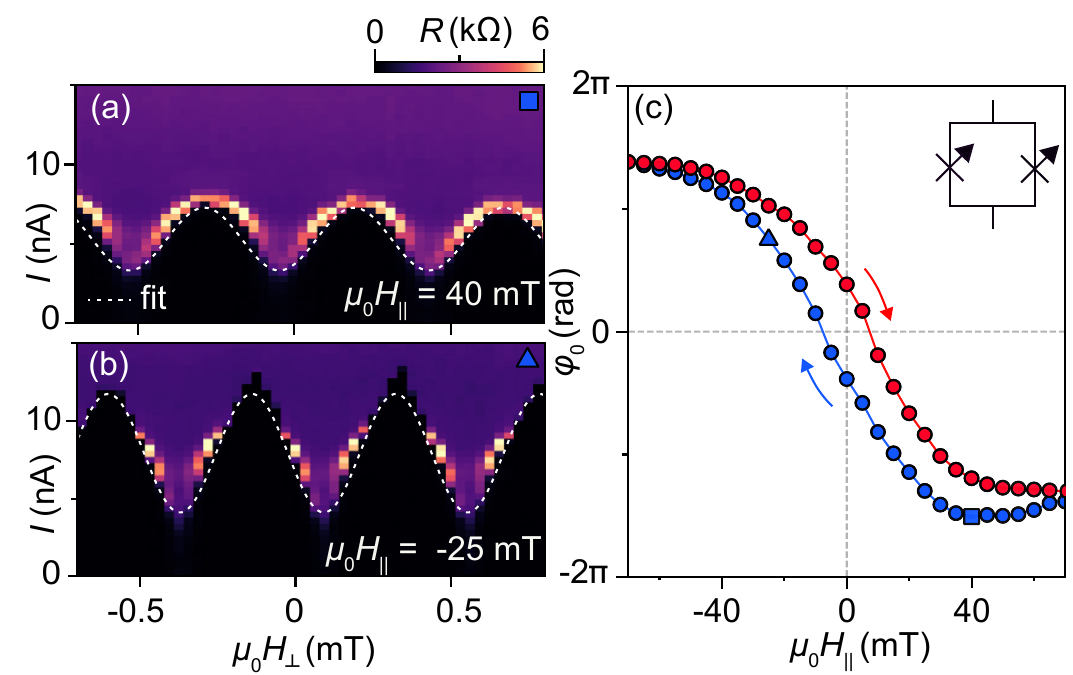}
    \caption{
    (a) and (b) Differential resistance, $R$, as a function of current bias, $I$, and perpendicular magnetic field, $\mu_0 H_\perp$, measured at fixed axial fields (a) $\mu_0 H_\parallel=40$~mT and (b) $-25$~mT.
    White dashed curves show sinusoidal fits to the switching-current boundaries.
    (c) Extracted relative phase offset, $\varphi_0$, as a function of $H_\parallel$ for opposite sweep directions.
    An arbitrary common phase offset is subtracted from both traces, such that only relative phase changes are determined.
    Data were taken at $V_{\rm BG}=10$~V and $V_{\rm PG}=5$~V.
     } 
    \label{fig:2} 
\end{figure}

To probe the anomalous phase shift of the target junction, we open the reference junction, tuning the interferometer to an asymmetric regime, with $I_{\rm SW}^{\rm R}>I_{\rm SW}^{\rm T}$, such that the flux-dependent modulation is predominantly sensitive to the target junction.
Applying a perpendicular magnetic field produces periodic oscillations of the switching current, with a period $\mu_0\,\delta H_\perp\approx 460~\mu$T, corresponding to one superconducting flux quantum through the interferometer loop.
Repeating these measurements at different axial magnetic fields reveals shifts of the interference pattern along the $H_\perp$ axis, as illustrated at $\mu_0 H_\parallel=40$ and $-25$~mT [Figs.~\ref{fig:2}(a) and \ref{fig:2}(b)].
We extract the relative phase offset by fitting the oscillations to $I_{\rm SW}(H_\perp)=A\sin(2\pi\,H_\perp/\delta H_\perp+\varphi_0)+I_0$.
As the axial field is reversed, the phase offset evolves continuously between opposite saturated values [Fig.~\ref{fig:2}(c)], in contrast to the abrupt $0$--$\pi$ transitions previously associated with discrete magnetic-domain reversals~\cite{Razmadze2023}.
The evolution is hysteretic, with distinct phase branches persisting at zero applied field, revealing a remanent phase bias.
Controlled excursions of the axial magnetic field further allow the remanent phase bias to be programmed at zero applied field, enabling operation as a Josephson phase battery; see the End Matter.

\begin{figure}[t]
    \includegraphics[width=0.48\textwidth]{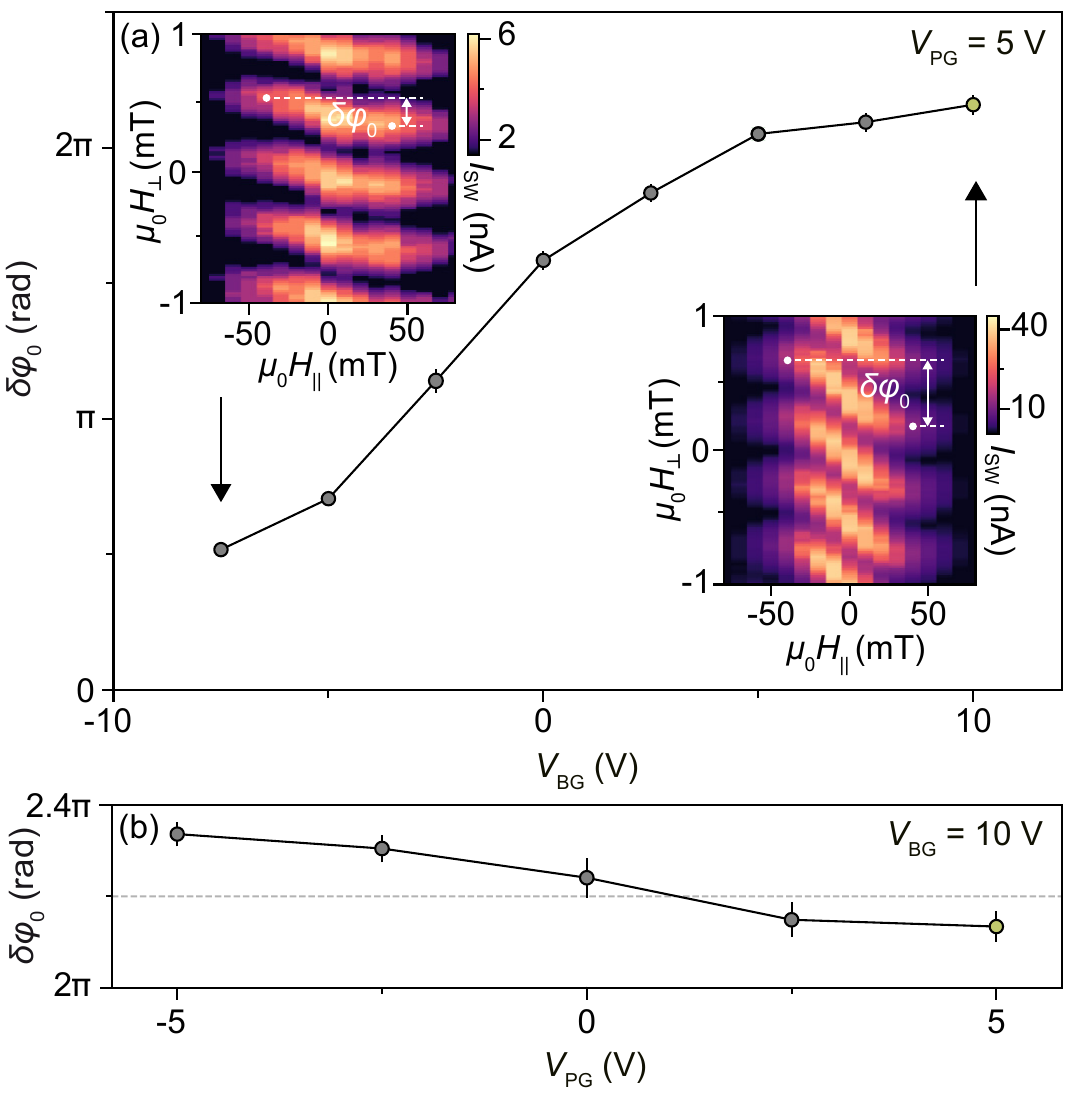}
    \caption{
    (a) Phase excursion, $\delta\varphi_0=\varphi_0(-40~{\rm mT})-\varphi_0(40~{\rm mT})$, as a function of global back-gate voltage, $V_{\rm BG}$, extracted along the same axial-field sweep direction.
    Insets show representative maps of switching-current, $I_{\rm SW}$, as a function of axial and perpendicular fields, $H_\parallel$ and $H_\perp$, at $V_{\rm BG}=-7$~V (upper left) and $10$~V (lower right), with $V_{\rm PG}=5$~V.
    (b) Corresponding phase excursion as a function of plunger-gate voltage, $V_{\rm PG}$, at fixed $V_{\rm BG}=10$~V.
    The target junction gate is adjusted to maintain an approximately constant switching-current ratio between the reference and target junctions.
    Error bars combine the propagated phase-fit uncertainties at $\mu_0 H_\parallel=\pm40$~mT with variations arising from the switching-current detection threshold.
    }
    \label{fig:3} 
\end{figure}

Having established the magnetic-history dependence of the anomalous phase, we next investigate its electrostatic control.
We repeat the interferometer measurements at different gate voltages, compensating with the target junction gate to maintain an approximately constant switching-current ratio $I_{\rm SW}^{\rm R}/I_{\rm SW}^{\rm T}$ and minimize changes in the interference response arising from junction asymmetry.
We also correct for the small geometric cross-coupling between the axial and perpendicular magnetic fields, as detailed in the Supplemental Material~\cite{SupplementalMaterial}.
To quantify the gate dependence, we define the phase excursion $\delta\varphi_0=\varphi_0(-40~{\rm mT})-\varphi_0(40~{\rm mT})$ between approximately saturated magnetic configurations, measured along the same axial-field sweep direction.
Taking the difference removes the arbitrary common phase offset, allowing us to quantify the phase response without determining the absolute Josephson phase.

Starting from the configuration used in Fig.~\ref{fig:2} ($V_{\rm BG}=10$~V, $V_{\rm PG}=5$~V), we decrease the back-gate voltage, progressively depleting carriers in the nanowire.
The phase excursion is strongly suppressed, from $\delta\varphi_0\approx2\pi$ at $V_{\rm BG}=10$~V to $\lesssim\pi/2$ at $V_{\rm BG}=-7$~V [Fig.~\ref{fig:3}(a)].
At more negative voltages, depletion of the junction prevents us from maintaining the same switching-current ratio through junction-gate compensation, although the observed trend suggests further suppression of the phase excursion.
If we instead fix $V_{\rm BG}$ to $10$~V and decrease $V_{\rm PG}$, which predominantly tunes the target-wire leads from the side, we observe a weaker but systematic enhancement of the phase excursion [Fig.~\ref{fig:3}(b)].

To further investigate the origin of the anomalous phase response, we examine the target junction independently, with the reference junction depleted.
At zero applied magnetic field, we observe nonreciprocal switching currents, yielding a superconducting diode efficiency of $-21\%$; see the End Matter.
In spin--orbit-coupled Josephson junctions, such nonreciprocity is closely linked to anomalous phase shifts, with higher harmonics of the current--phase relation providing the additional ingredient required for the diode effect~\cite{Maiani2023Nonsinusoidal, Reinhardt2024Link}.
Since the nonreciprocity is observed in the isolated target junction, it cannot arise from interference between two individually reciprocal junctions~\cite{SeoaneSouto2022Diode}, providing independent evidence for intrinsic symmetry breaking.
\begin{figure}[t]
    \includegraphics[width=0.48\textwidth]{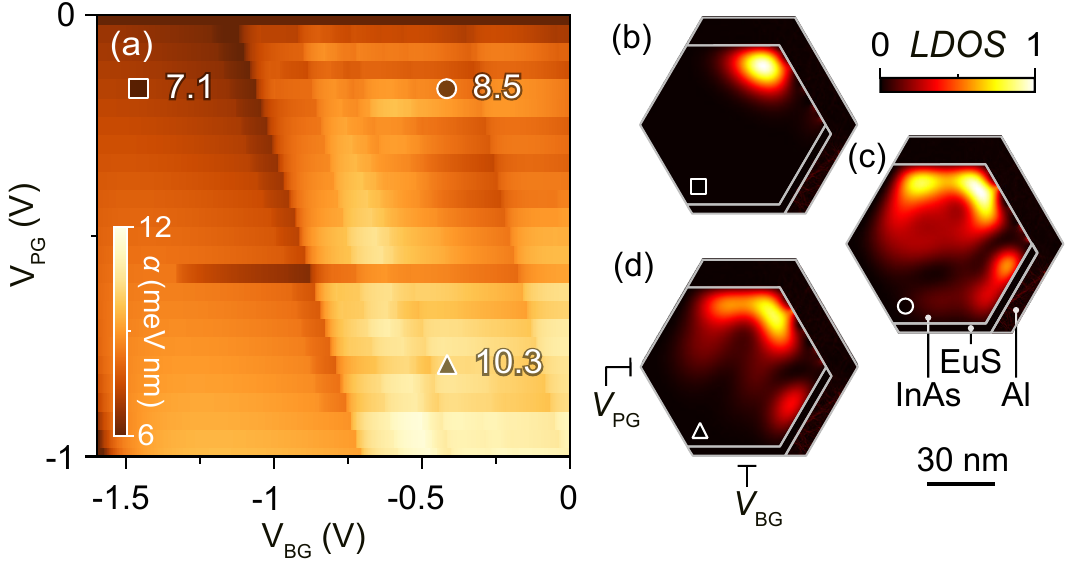}
    \caption{
    (a) Calculated effective Rashba spin--orbit coupling, $\alpha$, in the proximitized leads adjacent to the junction, as a function of back-gate voltage, $V_{\rm BG}$, and plunger-gate voltage, $V_{\rm PG}$, for the relative orientation of the Al and EuS shells and gate electrodes of the main device.
    The effective coupling is obtained by averaging the transverse Rashba components over the nanowire cross section, weighted by the local density of states (LDOS).
    (b)--(d) Normalized LDOS across the hybrid nanowire cross section for the three selected gate configurations indicated in (a).
    Increasing $V_{\rm BG}$ from (b) to (c) redistributes electronic states further into the semiconductor, whereas decreasing $V_{\rm PG}$ from (c) to (d) concentrates them near the corner where the InAs, EuS, and Al meet.
}
    \label{fig:4} 
\end{figure}

The contrasting responses to back-gate and plunger-gate tuning (Fig.~\ref{fig:3}) suggest that the anomalous phase depends not only on carrier density, but also on the spatial distribution of electronic states within the nanowire.
Because the Al and EuS shells cover selected facets, the two gate geometries can redistribute charge carriers differently across the nanowire cross section, modifying their effective spin--orbit coupling and magnetic proximity.
Microscopically, transverse Rashba spin--orbit coupling mixes spin-polarized subbands, making the Fermi velocity spin-dependent.
An exchange-field component along the spin–orbit axis then produces an electron–hole phase imbalance and hence an anomalous Josephson phase~\cite{Yokoyama2014Anomalous}.
Orbital effects arising from stray magnetic fields may also contribute to the phase shift~\cite{Liu2020, Davydova2022}.
To investigate the gate dependence of the phase response, we perform self-consistent microscopic electrostatic calculations of the hybrid nanowire, taking into account the shell orientations relative to the gate electrodes.

The simulations solve the coupled Schrödinger--Poisson equations, explicitly including the semiconductor, ferromagnetic insulator, and superconductor on equal footing, following the methods described in Refs.~\cite{Escribano2019, Escribano2021}.
The model yields the gate-dependent subband spectrum and spatially resolved density of states for a cross section corresponding to the Al-covered, proximitized nanowire segments adjacent to the junction.
By weighting the local contributions according to the spatial distribution of the occupied states, we extract effective parameters, including the superconducting gap, $\Delta$, exchange field, $h_{\rm ex}$, and Rashba spin--orbit coupling, $\alpha$.
Here, we focus on $\alpha$, while the gate dependences of $\Delta$ and $h_{\rm ex}$ are presented in the Supplemental Material~\cite{SupplementalMaterial}.
The simulated geometry is scaled down for numerical efficiency, and the calculated gate-voltage scales are not intended to match experiment quantitatively.
The calculated $\alpha$ exhibits contrasting responses to the two gate geometries: decreasing the back-gate voltage suppresses $\alpha$, whereas decreasing the plunger-gate voltage enhances it [Fig.~\ref{fig:4}(a)].

To understand these trends, we examine the spatial distribution of the occupied states at three representative gate configurations [Figs.~\ref{fig:4}(b)--(d)].
In our model, Rashba spin--orbit coupling arises from the local electric field in InAs and is set to zero in the Al and EuS shells.
The effective coupling therefore depends on both the local electric field and the degree of hybridization with the shells, which reduces the semiconductor weight of the electronic states.
Increasing the back-gate voltage shifts the states away from the Al shell and further into the semiconductor [Figs.~\ref{fig:4}(b) and \ref{fig:4}(c)], where the reduced hybridization enhances the effective $\alpha$.
Conversely, decreasing the plunger-gate voltage concentrates the states near the corner where InAs, EuS, and Al meet [Figs.~\ref{fig:4}(c) and \ref{fig:4}(d)].
In this case, stronger interfacial electric fields enhance $\alpha$, despite the increased proximity to the Al shell.
These contrasting responses depend on the orientation of the Al and EuS shells relative to the gate electrodes, which determines how the applied potentials redistribute electronic states within the nanowire.
Complementary Josephson calculations using an effective multichannel Rashba nanowire model show that the anomalous phase shift increases in magnitude with both spin--orbit coupling and nanowire width, highlighting its sensitivity to the transverse electronic structure; see the End Matter.
Together, the calculated gate dependences of $\alpha$ and their correspondence with the measured trends in $\delta\varphi_0$ suggest that electrostatically controlled spin--orbit coupling plays a central role in the anomalous Josephson phase response.

In conclusion, we demonstrate a continuously evolving anomalous Josephson phase in ferromagnetic InAs/EuS/Al nanowire junctions, extending phase control beyond discrete $0$--$\pi$ switching.
The phase response is hysteretic under axial magnetic-field sweeps and retains a programmable remanent bias at zero applied field, enabling operation as a Josephson phase battery.
Electrostatic gating strongly modifies the phase excursion, revealing its sensitivity to the transverse electronic structure of the hybrid nanowire.
The independent observation of a field-free Josephson diode effect provides additional evidence for intrinsic symmetry breaking in the ferromagnetic hybrid junction.
Self-consistent electrostatic simulations reveal a correspondence between the gate dependences of the effective Rashba spin--orbit coupling and the measured anomalous phase, supported by complementary Josephson calculations.
Together, these results establish ferromagnetic hybrid nanowires as a platform for magnetically programmable and electrostatically tunable Josephson phase biases.\\

\textit{Acknowledgments}---We thank C. Paya and P. San-Jose for valuable discussions. This research was supported by the Danish National Research Foundation; the European Innovation Council [Grant Agreement No.~101115548 (FERROMON)]; the European Research Council (Grant Agreement No.~856526); VILLUM FONDEN (Projects No.~43951 and 53097); the Spanish AEI through Grant No.~PID2024-161665NB-I00; and the Severo Ochoa Centers of Excellence program (CEX2024-001445-S), funded by MICIU/AEI/10.13039/501100011033, ``ERDF A way of making Europe'', and ``ESF+''.\\

\bibliography{Ref}

\vspace{3em}
\onecolumngrid
\begin{center}
    \textbf{\large End Matter}
\end{center}
\twocolumngrid

\begin{figure}[b]
    \includegraphics[width=0.48\textwidth]{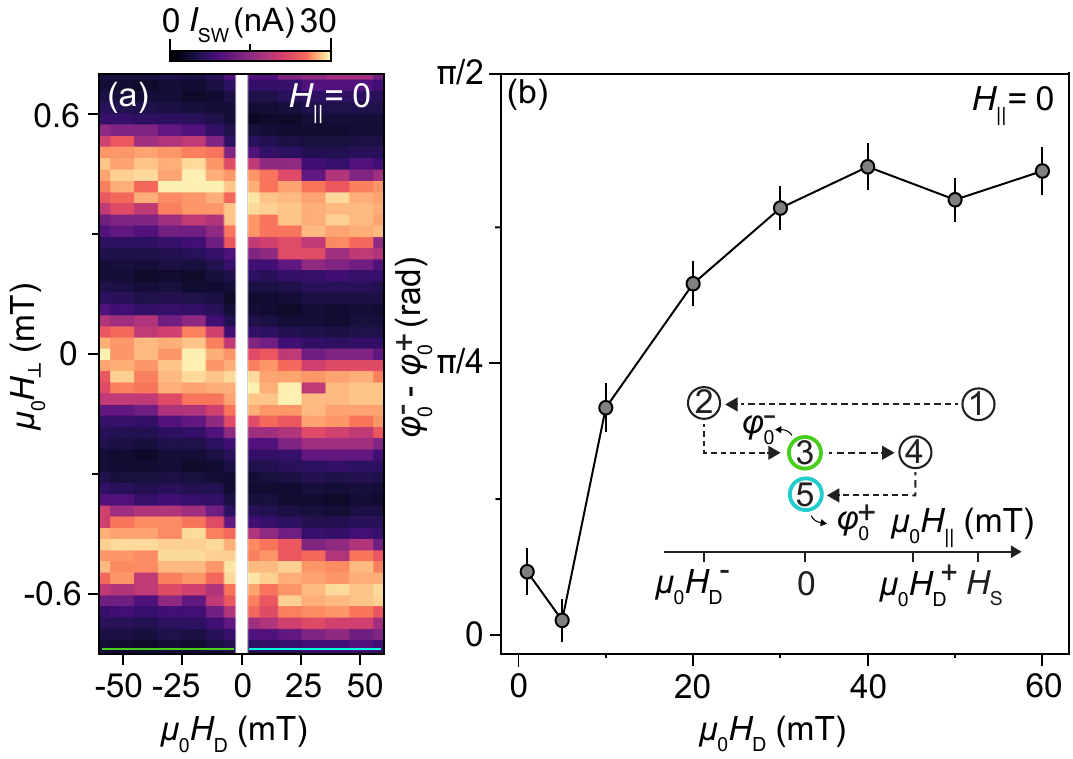}
    \caption{
    (a) Interferometer switching current, $I_{\rm SW}$, as a function of perpendicular magnetic field, $H_\perp$, and the preceding axial-field excursion, $H_{\rm D}$.
    Each interference pattern is measured at $H_\parallel=0$ after returning from the corresponding excursion.
    (b) Difference between the remanent phase offsets, $\varphi_0^--\varphi_0^+$, prepared by opposite magnetic-field histories, as a function of $\mu_0 H_{\rm D}$.
    Inset illustrates the magnetic-field protocol, consisting of initial saturation at $H_{\rm S} = 200$~mT, an excursion to $H_{\rm D}$, and subsequent zero-field readout.
     Data were taken at $V_{\rm BG}=10$~V and $V_{\rm PG}=5$~V.
    }
    \label{fig:5} 
\end{figure}

\textit{Appendix A: Josephson Phase Battery}---To investigate the magnetic programmability of the remanent phase bias, we measure the interferometer response following controlled axial magnetic-field excursions.
Starting from a saturated magnetic state at $\mu_0 H_{\rm S} = 200$~mT, the axial field is swept to a selected negative field, $H_{\rm D}^-$, and subsequently returned to $H_\parallel=0$, where the interference pattern is measured.
Without re-saturating the wire, we then continue to a positive field, $H_{\rm D}^+$, and return to zero for a second measurement.
Varying the magnitude of these field excursions produces systematic displacements of the interference patterns along $H_\perp$ [Fig.~\ref{fig:5}(a)].
We quantify this behavior by the difference between the remanent phase offsets following the negative and positive excursions, $\varphi_0^--\varphi_0^+$.
The phase difference evolves continuously with the magnitude of the field excursion, $\mu_0 H_{\rm D}$ [Fig.~\ref{fig:5}(b)], consistent with a gradual reconfiguration of the EuS magnetization and the exchange field experienced by the junction.
The magnetic-field excursion thus acts as a write operation, setting a remanent phase bias that persists after removal of the external field.
In this sense, the junction operates as a programmable Josephson phase battery.

\begin{figure}[b]
    \includegraphics[width=0.48\textwidth]{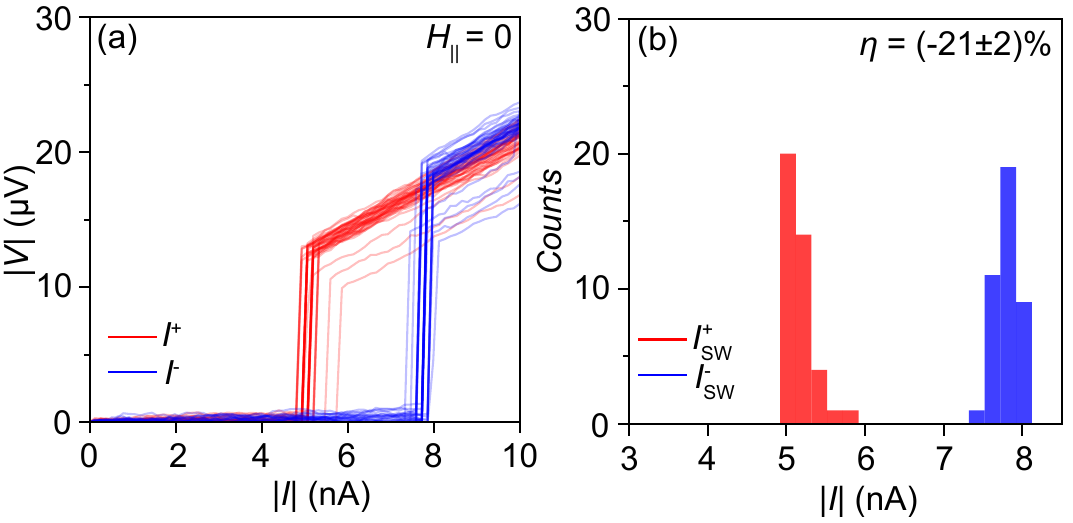}
    \caption{
    (a) Absolute dc voltage, $\vert V\vert$, as a function of current bias magnitude, $\vert I\vert$, for repeated sweeps in the positive and negative current directions at zero applied magnetic field.
    (b) Histograms of the positive and negative switching currents, $I_{\rm SW}^+$ and $|I_{\rm SW}^-|$, extracted from the measurements in (a).
    The distinct switching-current distributions yield a superconducting diode efficiency of $\eta=(-21\pm2)\%$. Data were taken at $V_{\rm BG}=10$~V and $V_{\rm PG}=5$~V.
  }
    \label{fig:6} 
\end{figure}

\textit{Appendix B: Josephson Diode Effect}---To characterize the nonreciprocal supercurrent response of the target junction, we deplete the reference junction and perform repeated dc current--voltage measurements at zero applied magnetic field, with fixed $V_{\rm BG}=10$~V and $V_{\rm PG}=5$~V [Fig.~\ref{fig:6}(a)].
To account for instrumental asymmetries, the measured current and voltage are corrected for systematic offsets introduced by the current source, transimpedance amplifier, and voltage preamplifier, following the procedure described in Ref.~\cite{Telkamp2025FieldFreeDiode}.
The positive switching current, $I_{\rm SW}^+$, is approximately 5~nA, whereas the magnitude of the negative switching current, $|I_{\rm SW}^-|$, reaches approximately 8~nA, resulting in a finite current range where the junction is superconducting for one bias direction and resistive for the other.
The corresponding switching-current histograms display clearly separated distributions for the two current directions [Fig.~\ref{fig:6}(b)].
Using the mean switching currents, we obtain a diode efficiency of $\eta=(I_{\rm SW}^+ -|I_{\rm SW}^-|)/( I_{\rm SW}^+ +|I_{\rm SW}^-|)=(-21\pm2)\%$.
The persistence of nonreciprocity in the isolated target junction supports an intrinsic Josephson diode effect, consistent with broken inversion symmetry of the hybrid structure and time-reversal symmetry breaking by the remanent EuS magnetization.

\textit{Appendix C: Effective model and calculation}---To investigate the dependence of the anomalous Josephson phase on spin--orbit coupling and transverse confinement, we model a Rashba semiconductor nanowire as a hard-wall strip of width $W$ subject to a Zeeman field, with a finite number of transverse modes, connected to semi-infinite superconducting leads described by Bogoliubov--de Gennes Hamiltonians.
\begin{figure}[t]
    \includegraphics[width=0.48\textwidth]{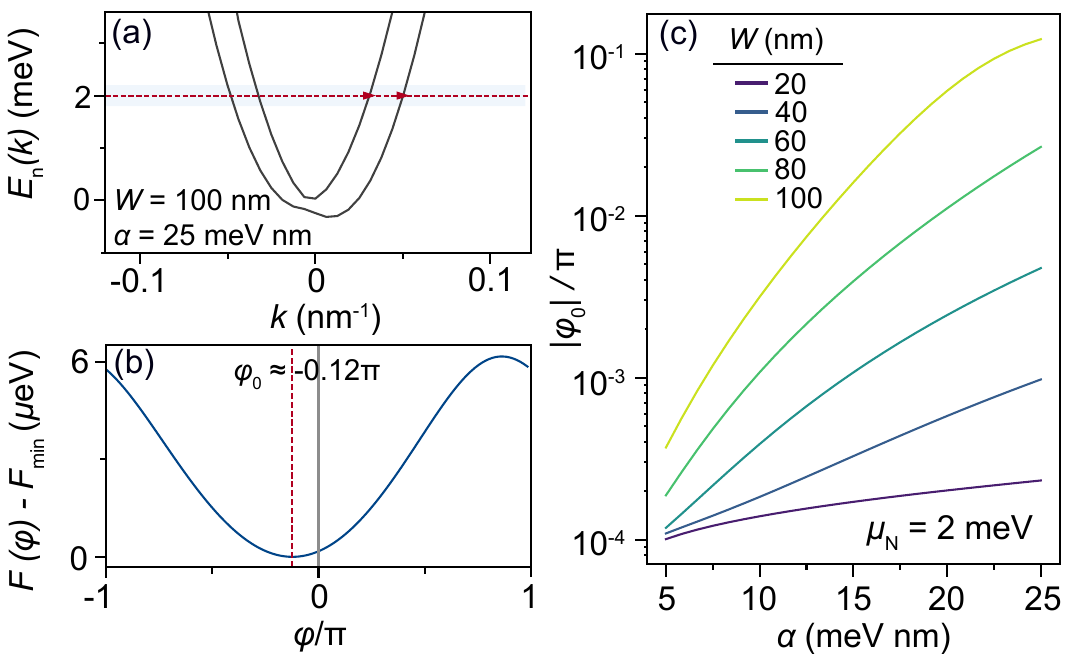}
    \caption{
    (a) Normal-state band structure of the two-mode Rashba nanowire model for $W=100$~nm, $\mu_N=2$~meV, and $\alpha=25$~meV\,nm, showing an asymmetric dispersion around $k=0$.
    (b) Corresponding free-energy difference, $F(\varphi)-F_{\rm min}$, as a function of superconducting phase difference across the junction, $\varphi$, exhibiting an anomalous minimum at $\varphi_0\approx-0.12\,\pi$.
    (c) Magnitude of the anomalous phase shift, $|\varphi_0|/\pi$, as a function of Rashba coupling, $\alpha$, at fixed $\mu_N=2$~meV and $\mu_S=10$~meV, showing an increase with both $\alpha$ and nanowire width, $W$.
    }
    \label{fig:7} 
\end{figure}

The transverse eigenfunctions $\chi_n(y)=\sqrt{2/W}\sin(n\pi y/W)$ have subband energies $\epsilon_n=\frac{\hbar^2}{2m^\ast}(\frac{\pi}{W})^2(n^2-1)$, where $m^\ast$ is the semiconductor effective mass and the lowest subband ($n=1$) defines the energy reference.
The chemical potentials in the normal segment, $\mu_{\rm N}$, and superconducting leads, $\mu_{\rm S}$, are measured relative to this reference.
Spin--orbit coupling is described by the Rashba Hamiltonian $H_{\rm R}=\alpha(k_x\sigma_y+k_y\sigma_x)$, where $\alpha$ is the Rashba coupling coefficient, $k_x$ and $k_y$ are the longitudinal and transverse wave vectors, and $\sigma_x$ and $\sigma_y$ are Pauli matrices acting on spin.
In the transverse basis, the momentum operator has matrix elements $P_{nm}=\langle n|-i\partial_y|m\rangle=-i\,4nm/[W(n^2-m^2)]$ for odd $n+m$, and zero otherwise.
The Rashba coupling therefore mixes different subbands through $\alpha P_{nm}\sigma_x$, while the longitudinal contribution is $\alpha k_x\sigma_y$.

Discretizing the longitudinal direction with spacing $a =1$~nm, and defining $q=k_xa$, $t_x=\hbar^2/(2m^\ast a^2)$, and $\alpha_x=\alpha/(2a)$, the normal-state electron Hamiltonian in region $r\in\{\rm N, S\}$ becomes
\begin{equation*}
    \begin{aligned}
      h^{(r)}_{nm}(q)
      &= \left[2t_x(1-\cos q)+\epsilon_n-\mu_r\right]
         \delta_{nm}\sigma_0 \\
      &\quad + 2\alpha_x\sin q\,\delta_{nm}\sigma_y \\
      &\quad + \alpha P_{nm}\sigma_x \\
      &\quad + \delta_{nm}{\bf B}_r\cdot\bm{\sigma}.
    \end{aligned}
\end{equation*}
Here, $\sigma_0$ and $\boldsymbol{\sigma}$ denote the identity and Pauli matrices in spin space, while $\mathbf B_r$ is the Zeeman-energy vector.
The Hamiltonian is extended to Bogoliubov--de Gennes space
in the Nambu basis
$\Psi=(\psi_\uparrow,\psi_\downarrow,\psi^\dagger_\downarrow,-\psi^\dagger_\uparrow)^{\rm T}$,
in which the spin-independent and Rashba terms of $h^{(r)}$ multiply $\tau_z$,
the Zeeman term $\mathbf{B}_r\cdot\boldsymbol{\sigma}$ multiplies $\tau_0$,
and the pairing term reads
$\Delta_r(\cos\varphi_r\,\tau_x+\sin\varphi_r\,\tau_y)\sigma_0$,
where $\boldsymbol{\tau}$ denotes Pauli matrices in particle--hole space.
The normal segment has $\Delta_{\rm N}=0$, while the superconducting leads have $\Delta_{\rm S}=\Delta$, with phases $\varphi_{\rm L}=\varphi$ and $\varphi_{\rm R}=0$.

We calculate the equilibrium energy--phase relation using an embedded Matsubara formulation, representing the semi-infinite superconducting leads by their exact surface self-energies.
The left-lead self-energy at phase $\varphi$ is obtained from its zero-phase value by a Nambu gauge rotation, $\Sigma_{\rm L}(i\omega,\varphi)=U(\varphi)\Sigma_{\rm L}(i\omega,0)U^\dagger(\varphi)$, with $U(\varphi)=\exp(-i\varphi\tau_z/2)$. 
Recursively eliminating the normal segment yields a phase-dependent boundary matrix, $M(i\omega,\varphi)$, from which the equilibrium free-energy difference is calculated as 
\begin{equation*}
    \begin{aligned}
      F(\varphi)-F(0)
        &= -\frac{1}{2\pi}\int_0^\infty d\omega \\
        &\quad\times \operatorname{Re}\log\det\!\left[
          M(i\omega,\varphi)M^{-1}(i\omega,0)
          \right].
    \end{aligned}
\end{equation*}
The anomalous phase is obtained by minimizing the resulting free energy, $\varphi_0=\operatorname*{arg\,min}_{\varphi\in[-\pi,\pi)}F(\varphi)$, using local periodic interpolation.
Numerical integration and phase-reconstruction details are provided in the Supplemental Material~\cite{SupplementalMaterial}.

To investigate the low-density regime relevant to the experiment, we retain two transverse modes to account for intersubband mixing, while keeping $\mu_{\rm N}$ below the second subband threshold.
As a representative example, we use $W=100$~nm, $\Delta=0.2$~meV, $\mu_{\rm S}=10$~meV, $\mu_{\rm N}=2$~meV, $\alpha=25$~meV\,nm, junction length $L_{\rm N}=100$~nm, and equal Zeeman-energy vectors $\mathbf B_{\rm N}=\mathbf B_{\rm S}=(0.1,0.1,0)$~meV.
The resulting band structure and energy--phase relation are shown in Figs.~\ref{fig:7}(a) and \ref{fig:7}(b).
In this model, a finite exchange-field component along $y$, parallel to the effective Rashba field associated with longitudinal motion, is required to generate an anomalous phase shift.
Such a component could arise from local noncollinearity of the EuS magnetization~\cite{Nandi2026, Telkamp2025FieldFreeDiode}.
The higher chemical potential in the superconducting leads accommodates an additional transverse mode, introducing a band mismatch that substantially influences the anomalous phase.
Setting $\mu_{\rm S}=\mu_{\rm N}$ reduces $\varphi_0$ from approximately $-0.12\,\pi$ to $-0.02\,\pi$.
The anomalous phase is therefore sensitive to the electronic structure of the superconducting leads, not only to that of the weak link, consistent with the gate-tuned lead simulations in Fig.~\ref{fig:4}.
Keeping $\mu_{\rm S}=10$~meV and $\mu_{\rm N}=2$~meV, the magnitude of $\varphi_0$ increases systematically with both Rashba coupling and nanowire width [Fig.~\ref{fig:7}(c)], demonstrating the sensitivity of the anomalous Josephson phase to spin--orbit coupling and transverse confinement.

\end{document}


\title{Supplemental Material:\\ Electrostatic and magnetic control of the anomalous Josephson phase in\\ ferromagnetic hybrid junctions}

\author{D. Rana}
\affiliation{Niels Bohr Institute, University of Copenhagen, 2100 Copenhagen, Denmark}
\author{O. S. Kurapati}
\affiliation{Niels Bohr Institute, University of Copenhagen, 2100 Copenhagen, Denmark}
\author{N. Vanjarapu}
\affiliation{Niels Bohr Institute, University of Copenhagen, 2100 Copenhagen, Denmark}
\author{J. Zhao}
\affiliation{Niels Bohr Institute, University of Copenhagen, 2100 Copenhagen, Denmark}
\author{J. B. F. Wolters}
\affiliation{Niels Bohr Institute, University of Copenhagen, 2100 Copenhagen, Denmark}
\author{Y. Liu}
\affiliation{Niels Bohr Institute, University of Copenhagen, 2100 Copenhagen, Denmark}
\author{P.~Krogstrup}
\affiliation{Niels Bohr Institute, University of Copenhagen, 2100 Copenhagen, Denmark}
\author{A. L. R. Manesco}
\affiliation{Niels Bohr Institute, University of Copenhagen, 2100 Copenhagen, Denmark}
\author{S. D. Escribano}
\affiliation{Department of Condensed Matter Physics, Weizmann Institute of Science, Rehovot 7610001, Israel}
\author{E. Prada}
\affiliation{Instituto de Ciencia de Materiales de Madrid (ICMM), CSIC, 28049 Madrid, Spain}
\author{S. Vaitiek\.{e}nas}
\affiliation{Niels Bohr Institute, University of Copenhagen, 2100 Copenhagen, Denmark}

\date{\today}

\maketitle

\section*{Sample Preparation}

InAs nanowires were grown on InAs(111)B substrates (Semiconductor Wafer Inc.) by Au-catalyzed vapor--liquid--solid growth in a solid-source molecular beam epitaxy (MBE) system (Varian GEN-II) at a background pressure below $10^{-10}$ Torr. 
Following oxide desorption at $540^{\circ}\,\mathrm{C}$ under $\mathrm{As}_2$, growth proceeded at $420^{\circ}\,\mathrm{C}$ for 2100~s with an $\mathrm{As}_2$/In ratio of 11, yielding nanowires approximately 100~nm in diameter and $10~\mu$m long. 
The substrates were then transferred under vacuum to a metal deposition chamber, where nominally 2~nm of EuS layer was deposited on two facets at $180^{\circ}\,\mathrm{C}$ and 0.01~nm/s, with a substrate tilt of approximately $45^{\circ}$ towards the source. 
Al was subsequently deposited below $-30^{\circ}\,\mathrm{C}$ at 0.05~nm/s, with a beam--substrate angle of $27^{\circ}$. Substrate rotation enabled coverage of three facets, with an Al thickness of $\sim$20~nm on the central facet.

Devices were fabricated using standard electron-beam lithography (Elionix ELS-F125, 125 keV). 
Josephson junctions were patterned using an adhesion promoter (Allresist, AR 300-80) and 13\% CSAR 62 resist (Allresist, AR-P 6200), followed by selective wet etching of the \textit{in situ} Al shell. 
The etchant was a 10~mN TMAH solution prepared by mixing 29.9~g of IPA with 2.5~mL of 0.17~N TMAH (Allresist, AR 300-475). 
Etching was carried out at $50^{\circ}\,\mathrm{C}$ for 23~s, under conditions optimized to minimize the impact on the EuS layer. Following native oxide removal by Ar-ion plasma cleaning (rf ion source, 25~W, 18~mTorr, 9.5~min), 200~nm thick \textit{ex-situ} Al leads were deposited by electron-beam evaporation (AJA International Inc., Orion). 
Junction gates consisting of Ti/Au (5/150~nm) were separated from the nanowire by a 13~nm thick $\mathrm{HfO}_x$ dielectric grown by atomic layer deposition (Veeco, Savannah S100).

\begin{figure}[b!]
    \includegraphics[width=0.48\textwidth]{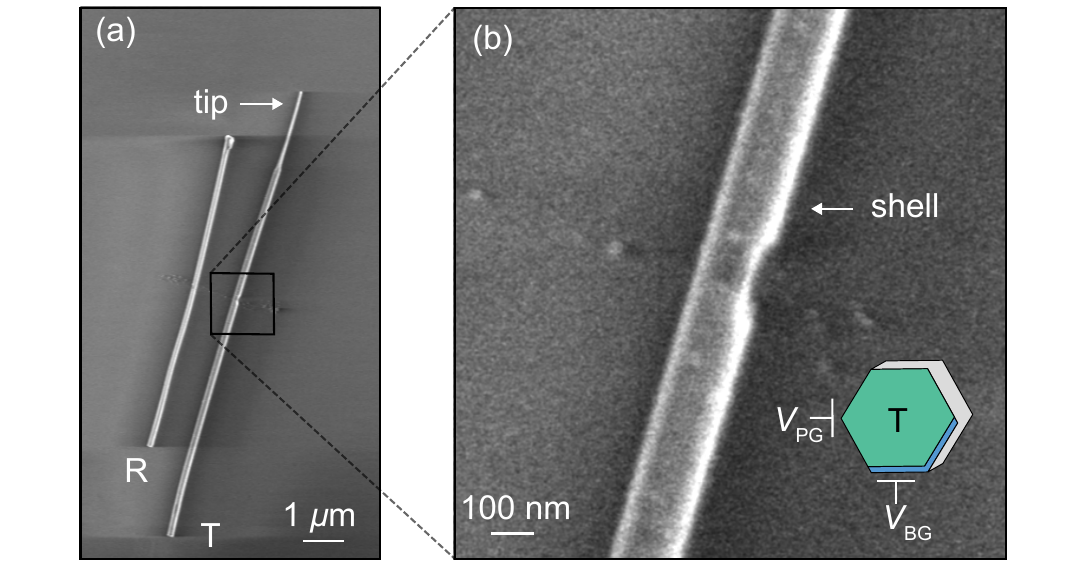}
    \caption{
    (a) Electron micrograph of the reference (R) and target (T) nanowires after junction etching, before fabrication of the gates and Al leads.
    The tip of the target wire is indicated.
    (b) Magnified view of the target wire junction region, showing the shell-covered segments and the approximately 100~nm long junction.
    The Al shell was removed in this region while retaining the EuS layer.
    Inset: inferred orientation of the Al and EuS shells relative to the back and plunger gates.
    }
    \label{fig:S1} 
\end{figure}

\section*{Measurements}

Transport measurements were performed using dc and low-frequency ac lock-in (Stanford Research, SR830) techniques in a cryofree dilution refrigerator (Oxford Instruments, Triton 400) equipped with a three-axis (1,~1,~6)~T vector magnet and a base temperature of 20~mK. 
All measurement lines were filtered at cryogenic temperatures using rf and RC filters with a cutoff frequency of  65~kHz (QDevil, QFilter). 
Both ac and dc currents were transamplified using an $I$--$V$ converter (Basel Precision Instruments, SP983) with a gain of $10^{6}$, and the voltages were amplified using low-noise preamplifiers (Stanford Research, SR560) with a gain of $10^{3}$. 
Four-probe differential resistance measurements were performed using an ac current excitation of 0.5~nA at 19.5~Hz. 
The in-plane magnetic field was aligned with the nanowire axis, accounting for the wire orientation relative to the chip axes. Before extracting the phase from the interferometer oscillations, the perpendicular-field axis was corrected for geometric cross-talk arising from an approximately $1^{\circ}$ out-of-plane misalignment between the chip and the vector-magnet coordinate system.

\section*{Shell Orientation}

We infer the orientation of the Al and EuS shells in the main device by examining the nanowire tip and the shell profile visible in electron micrographs taken after junction etching, but before fabrication of the gates and Al leads [Fig.~\ref{fig:S1}(a)].
A magnified view of the target wire [Fig.~\ref{fig:S1}(b)] resolves the approximately 100~nm long etched region, where the Al shell was selectively removed while retaining the EuS layer, and the shell-covered segments on either side.
Combining the visible morphology with the known facet coverage during growth, we infer the cross-sectional orientation of the target wire relative to the back and plunger gates, as illustrated in the inset of Fig.~\ref{fig:S1}(b).

The orientation of the shell-covered facets relative to the gate electrodes depends on the rotation of the nanowire about its axis during transfer.
Consequently, devices with nominally identical lithographic layouts may have different relative orientations of the Al and EuS shells.
Since the back gate acts from below and the plunger gates from the side and above, their potentials redistribute electronic states differently across the nanowire cross section, modifying the effective spin--orbit coupling and overlap with regions of induced exchange and superconducting pairing.
The inferred orientation is used in the electrostatic calculations discussed in the main text.

\begin{figure}[t!]
    \includegraphics[width=0.48\textwidth]{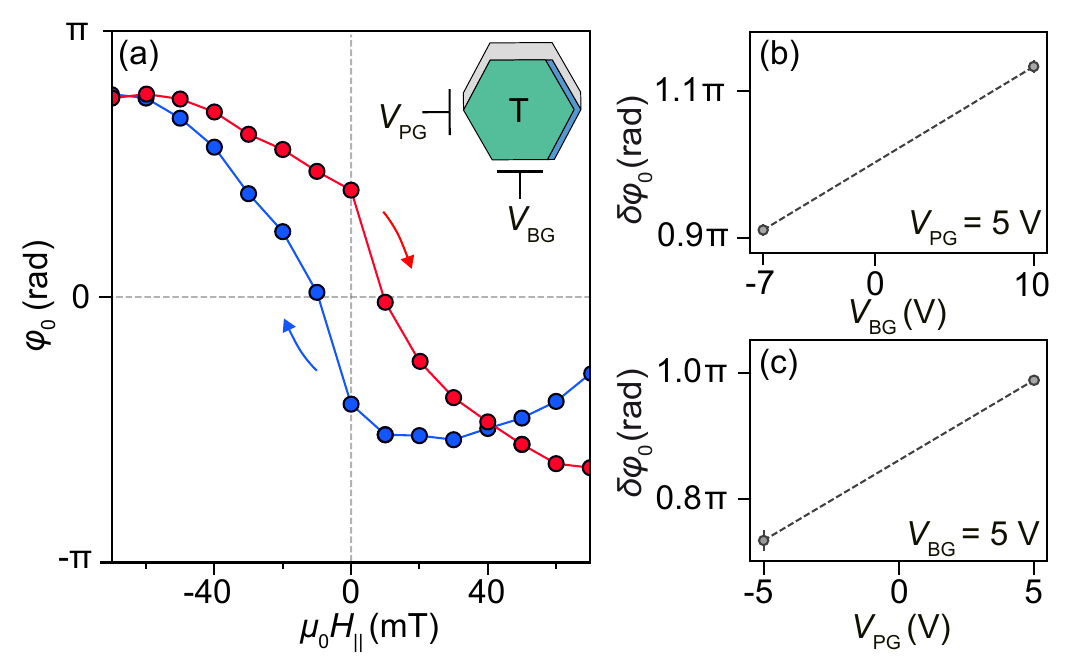}
    \caption{
    (a) Phase offset, $\varphi_0$, extracted from interferometer measurements on device 2 as a function of axial magnetic field, $H_\parallel$, for increasing (red) and decreasing (blue) field sweep directions.
    The branches remain separated at zero applied field.
    Inset: schematic cross section of the target wire, illustrating the independently determined shell orientation.
    Data were taken at $V_{\rm BG}=10$~V and $V_{\rm PG}=5$~V.
    (b) Phase excursion, $\delta\varphi_0$, as a function of $V_{\rm BG}$ at fixed $V_{\rm PG}=5$~V.
    (c) Corresponding phase excursion as a function of $V_{\rm PG}$ at fixed $V_{\rm BG}=5$~V.
    Connecting dashed lines are guides to the eye.
    }
    \label{fig:S2} 
\end{figure}

\section*{Anomalous Phase Response in Device 2}

To assess the reproducibility of the anomalous phase response, we performed interference measurements on a second device (device 2) with nominally the same lithographic design as the main device (device 1).
Device 2 exhibits a continuous, hysteretic evolution of $\varphi_0$ under axial magnetic-field sweeps [Fig.~\ref{fig:S2}(a)]. 
The two sweep directions follow distinct branches that remain separated at zero applied field, reproducing the history-dependent phase bias observed in device 1.
Electrostatic tuning also modifies the phase excursion $\delta\varphi_0$. 
Decreasing $V_{\mathrm{BG}}$ from $10$ to $-7$~V reduces $\delta\varphi_0$ [Fig.~\ref{fig:S2}(b)], consistent with the back-gate response of device 1. 
Decreasing $V_{\mathrm{PG}}$ from $5$ to $-5$ V likewise reduces $\delta\varphi_0$ [Fig.~\ref{fig:S2}(c)], opposite to the trend observed in device 1. 
Thus, both devices exhibit electrostatic control of the anomalous phase, although their plunger-gate responses differ.

The contrasting plunger-gate responses can be understood in terms of the different orientation of the shell-covered facets relative to the gates, despite the nominally identical lithographic design.
For the independently determined orientation illustrated in the inset of Fig.~\ref{fig:S2}(a), decreasing $V_{\mathrm{BG}}$ is expected to redistribute electronic states toward the upper-left Al-covered region and away from the EuS-covered side. 
Conversely, decreasing $V_{\mathrm{PG}}$ at positive $V_{\mathrm{BG}}$ is expected to favor the lower-right EuS-covered region over the upper Al-covered facets.
The two gate geometries therefore modify the overlap with the magnetic and superconducting regions differently than in device 1, providing a qualitative explanation for distinct plunger-gate dependences of $\delta\varphi_0$.

\begin{figure}[t!]
    \includegraphics[width=0.48\textwidth]{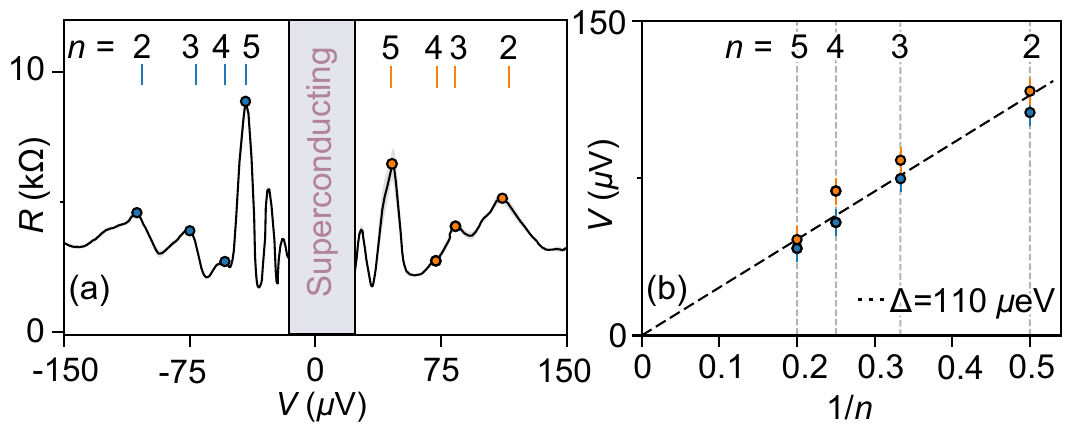}
    \caption{
    (a) Differential resistance $R$ versus measured voltage drop $V$,
averaged over five axial-field values centered on zero
from Fig. 1(c) of the main text.
Coloured markers identify the peaks assigned to MAR
orders $n=2$--$5$; shading indicates the superconducting branch.
(b) Absolute peak voltages versus $1/n$ for the two voltage
polarities. The dashed line is a fit to
$|V_n|=2\Delta/(en)$, yielding
$\Delta \simeq (110 \pm 3)\,\mu\mathrm{eV}$.
}
    \label{fig:S3} 
\end{figure}

\begin{figure}[t!]
    \includegraphics[width=0.51\textwidth]{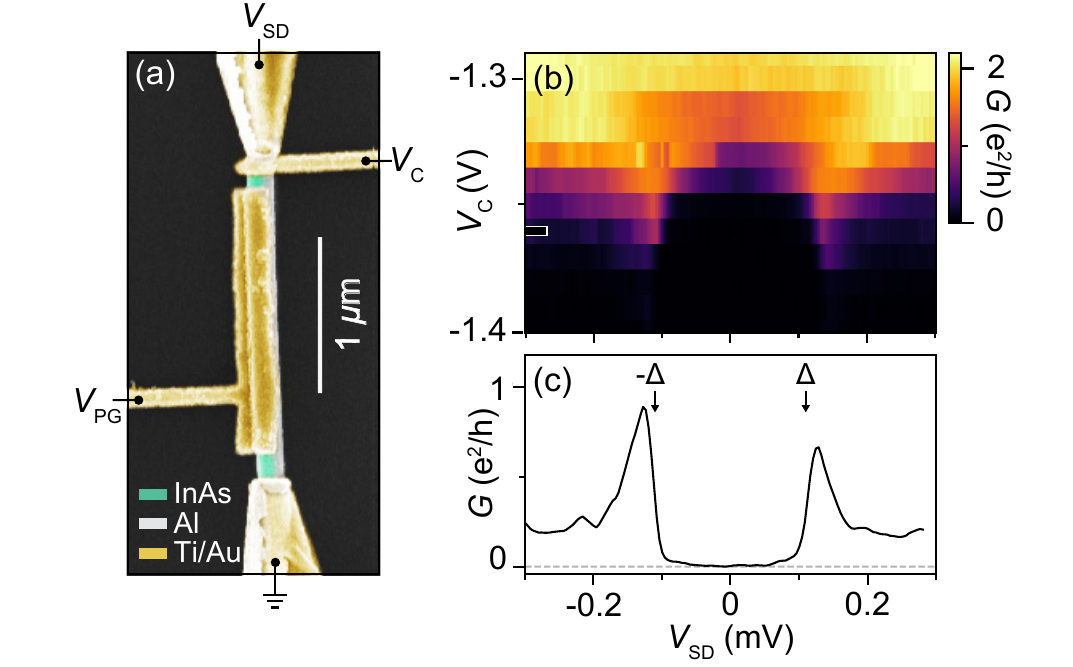}
    \caption{  
    (a) Electron micrograph of the tunneling device, with the Al shell removed from the upper half of the nanowire, which is contacted by a normal-metal lead, and a short junction covered by a top tunnel gate.
    (b) Differential conductance, $G=dI/dV_{\rm SD}$, as a function of source--drain bias, $V_{\rm SD}$, and local gate voltage, $V_{\rm C}$.
    (c) Conductance line cut at the gate voltage marked in (b), showing suppressed subgap conductance and peaks near $V_{\rm SD}=\pm120~\mu\mathrm{V}$.
    Markers at $\pm\Delta$ indicate the gap scale extracted from the MAR analysis in Fig.~\ref{fig:S3}.
    Data were taken at $V_{\rm PG}=0$.
    }
    \label{fig:S4} 
\end{figure}

\section{Subgap Transport and Spectroscopy}

To characterize finite-bias transport within the superconducting field window, we analyze differential-resistance traces from the increasing axial-field sweep in Fig.~1(c) of the main text.
We average traces at five axial-field values near zero, selected to have comparable switching currents, to reduce noise.
Plotting the averaged differential resistance, $R(I)$, against the measured dc voltage, $V(I)$, reveals peaks at both voltage polarities outside the zero-voltage superconducting branch [Fig.~\ref{fig:S3}(a)].
We assign the indicated peaks to multiple Andreev reflections (MARs) of orders $n=2$--$5$, with expected subharmonic voltages $e|V_n|=2\Delta/n$.
The peak positions scale approximately linearly with $1/n$ [Fig.~\ref{fig:S3}(b)], yielding an effective superconducting gap of $\Delta=(110\pm3)~\mu$eV.
Additional peaks deviating from the MAR sequence may originate from subgap structure in the leads or localized states within the junction~\cite{Abay2014,Bezuglyi2017,Ridderbos2019}.

Complementary voltage-biased, $V_{\rm SD}$, spectroscopy was performed on a tunneling device with a normal lead, fabricated from a nanowire from the same growth batch [Fig.~\ref{fig:S4}(a)].
The tunnel barrier was tuned using the local gate voltage $V_{\rm C}$.
Decreasing $V_{\rm C}$ suppressed the low-bias differential conductance, $G=dI/dV_{\rm SD}$, indicating a transition into the tunneling regime [Fig.~\ref{fig:S4}(b)].
A line cut in the regime with $G<e^2/h$ reveals a superconducting gap with conductance peaks near $V_{\rm SD}=\pm120~\mu$V [Fig.~\ref{fig:S4}(c)].
The corresponding gap scale agrees approximately with the independent MAR estimate, indicated by the markers at $eV_{\rm SD}=\pm\Delta$.

\begin{figure*}[t]
    \includegraphics[width=\textwidth]{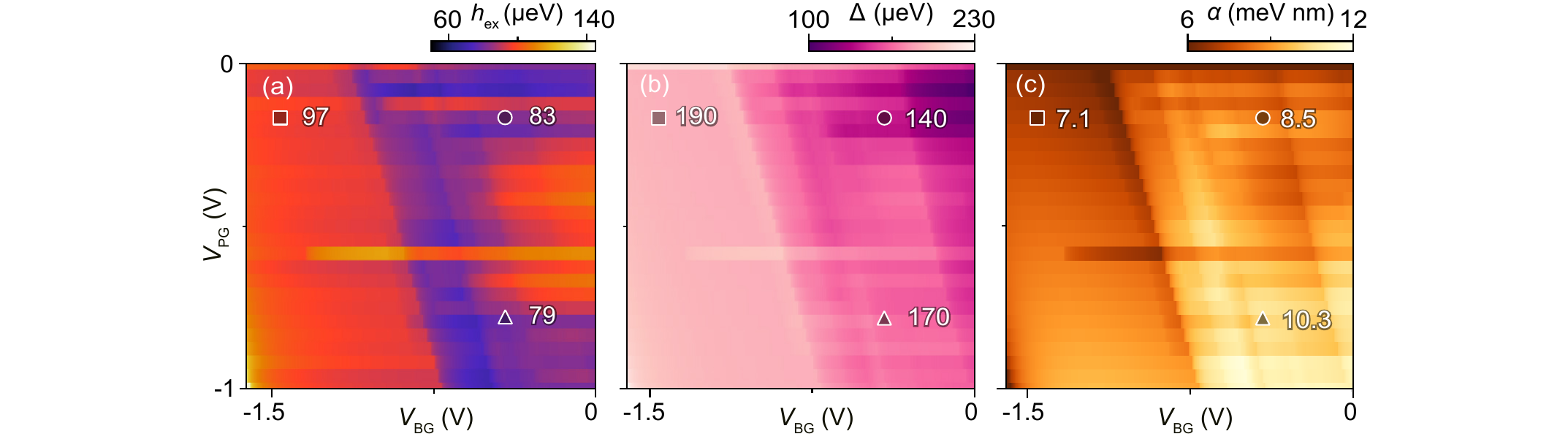}
    
    \caption{(a)-(c) Calculated exchange field $h_{\rm ex}$, induced superconducting gap $\Delta$ and effective Rashba spin–orbit coupling as a function of back gate
    $V_{\rm{BG}}$ and plunger gate $V_{\rm{PG}}$ voltages.
    }
    \label{fig:S5} 
\end{figure*}

\section{Self-consistent electrostatic simulations}
We model the electronic structure of the hybrid nanowire cross section in the proximitized leads adjacent to the junction. 
The electrostatic potential is obtained by solving the Poisson equation self-consistently, treating the mobile charge within the Thomas–Fermi approximation. 
The resulting potential is then incorporated into a Bogoliubov–de Gennes Hamiltonian, in which the semiconducting InAs core, ferromagnetic insulator EuS layer, and superconducting Al shell are treated on equal footing. 
This approach yields the low-energy spectrum and local density of states (LDOS) of the occupied subbands.
Following Ref.~\cite{Escribano2019Electrostatic}, we use the Hamiltonian
\begin{align}
H ={}& \Big[\vec{k}^{T}\frac{\hbar^{2}}{2m^{*}(\vec r)}\vec{k}
      + E_{\rm F}(\vec r) - e\phi(\vec r) \nonumber\\
 &+ \tfrac{1}{2}\big(\vec{\alpha}_{\rm R}(\vec r)\cdot(\vec\sigma\times\vec k)
      + (\vec\sigma\times\vec k)\cdot\vec{\alpha}_{\rm R}(\vec r)\big)\Big]\tau_z \nonumber\\
 &+ h_{\rm ex}(\vec r)\,\sigma_x\tau_0
  + \Delta(\vec r)\,\sigma_0\tau_x ,
\label{eq:S1}
\end{align}
written in the Nambu basis $\Psi(\vec r)=\big(\psi_\uparrow(\vec r),\psi_\downarrow(\vec r),\psi^\dagger_\downarrow(\vec r),-\psi^\dagger_\uparrow(\vec r)\big)^{T}$, where $\sigma_i$ and $\tau_i$ are Pauli matrices acting in spin and particle–hole space, respectively. 
We focus on the bulk properties of the hybrid nanowire and therefore assume translational invariance along the nanowire axis, taken as the $x$ direction, and solve the remaining problem in the transverse $(y,z)$ cross-section, with
\begin{equation}
\vec{r}=(y,z), \qquad
\vec{k}=(k_x,-i\partial_y,-i\partial_z),
\end{equation}
where $k_x$ is a good quantum number. 
The effective mass, $m^*$, conduction-band bottom, $E_{\rm F}$, exchange field, $h_{\rm ex}$, and superconducting pairing potential, $\Delta$, are modeled as piecewise-constant parameters across the different material regions. Here $E_{\rm F}(\vec r)$ is the conduction-band bottom of each material relative to the Fermi level, and thus encodes the band offsets at the
InAs/EuS and InAs/Al interfaces.

The Rashba spin-orbit coupling is included only in the semiconductor and is determined by the local electrostatic field. Following Ref.~\cite{Escribano2020Rashba}, we write
\begin{equation}
\vec{\alpha}_{\rm R}(\vec{r})
=
\frac{eP_{\rm fit}^2}{3}
\left[
\frac{1}{\Delta_{\rm g}^2}
-
\frac{1}{(\Delta_{\rm g}+\Delta_{\rm so})^2}
\right]
\vec{\nabla}\phi(\vec{r}),
\end{equation}
where $\Delta_{\rm g}$ is the semiconductor band gap, $\Delta_{\rm so}$ is the split-off gap, and $P_{\rm fit}$ is an effective Kane coupling accounting for the semiconductor material and crystal properties. 
The material parameters used in the simulations are summarized
in Table~\ref{Table:parameters}, with the values of $P_{\rm fit}$, $\Delta_{\rm g}$, and $\Delta_{\rm so}$ given in Ref.~\cite{Escribano2020Rashba}.
We also include disorder at the outer
surface of the superconducting shell to account for interface
inhomogeneity.

The electrostatic potential entering the Hamiltonian is computed
self-consistently. 
The calculation includes the full transverse geometry of the hybrid nanowire and its electrostatic environment, including the back and the plunger gates. 
At each gate configuration, the potential satisfies
\begin{equation}
    \vec{\nabla}\cdot
    \left[\epsilon(\vec{r})\vec{\nabla}\phi(\vec{r}) \right] =
    -\rho_{\rm surf}(\vec{r})-\rho_{\rm mobile}(\vec{r}),
\end{equation}
where $\epsilon(\vec{r})$ is the position-dependent dielectric constant and $\rho_{\rm surf}(\vec{r})$ describes surface charge accumulation typically present on the uncovered facets of InAs nanowires. 
The mobile charge density is approximated using the Thomas-Fermi
expression
\begin{equation}
    \begin{aligned}
        \rho_{\rm mobile}(\vec{r})
        &\simeq \rho_{\rm mobile}^{\rm (TF)}(\vec{r})\\
        &=-\frac{2e\sqrt{2}}{3\pi^2\hbar^3}
        \left[
        m^*|e\phi(\vec{r})-E_{\rm F}|\right]^{3/2}
        \\
        &
        \times f[-(e\phi(\vec{r})-E_{\rm F})],
    \end{aligned}
\end{equation}
where $f$ is the Fermi function. 
This approximation captures the screening from the occupied semiconductor states while keeping the electrostatic calculation
computationally tractable. 
The Poisson and Thomas--Fermi equations are solved self-consistently following the procedure of Ref.~\cite{Escribano2019Electrostatic}, including the surface-charge layer and the band offset $V_{\rm Al}$ at the InAs/Al interface, using the numerical routines of Ref.~\cite{Escribano2020Majorana} and the parameters in Table~\ref{Table:parameters}.

The calculated gate dependences of $h_{\mathrm{ex}}$, $\Delta$, and $\alpha$ are presented in Fig.~\ref{fig:S5}.
Increasing $V_{\mathrm{BG}}$ reduces the calculated
$h_{\rm ex}$ [Fig.~\ref{fig:S5}(a)], opposite to the measured trend in $\delta\varphi_0$. 
In contrast, the effective Rashba coupling increases
[Fig.~\ref{fig:S5}(c)], supporting its central role
in the observed gate dependence of the anomalous phase.

\begin{table}[t]
\caption{Parameters used for the simulations. The temperature is fixed to $10~\mathrm{mK}$ in all simulations.}
\label{Table:parameters}

\renewcommand{\arraystretch}{1.15}

\begin{ruledtabular}
\begin{tabular}{lccc}

\textbf{Material} & \textbf{Parameter} & \textbf{Value} & \textbf{Refs.} \\

\hline

InAs
& width
& $80~\mathrm{nm}$
& -- \\

& $m_{\mathrm{eff}}$
& $0.023m_0$
& \cite{Vurgaftman2001BandParameters} \\

& $E_{\mathrm{F}}$
& $0$
& \cite{Liu2020InAsEuSInterfaces} \\

& $h_{\mathrm{ex}}$
& $0$
& -- \\

& $\Delta$
& $0$
& -- \\

& $\alpha$
& \multicolumn{2}{c}{see text for details} \\

& $\epsilon_{\mathrm{InAs}}$
& $15.5\epsilon_0$
& \cite{Levinshtein1996Handbook} \\

& $\rho_{\mathrm{surf}}$
& $2\times10^{-3}~e/\mathrm{nm}^{3}$
& \cite{Thelander2010InAsNanowires, Reiner2020Spectroscopic} \\

\hline

Al
& width
& $10~\mathrm{nm}$
& -- \\

& oxidation width
& $2~\mathrm{nm}$
& -- \\

& $m_{\mathrm{eff}}$
& $m_0$
& \cite{Segall1961Aluminum} \\

& $E_{\mathrm{F}}$
& $-11.7~\mathrm{eV}$
& \cite{AshcroftMermin1976} \\

& $h_{\mathrm{ex}}$
& $0$
& -- \\

& $\Delta$
& $0.23~\mathrm{meV}$
& \cite{Chang2015HardGap} \\

& $\alpha$
& $0$
& \cite{Chiang2017SpinHall} \\

& $\epsilon_{\mathrm{Al}}$
& $\infty$
& -- \\

& $V_{\mathrm{Al}}$
& $0.2~\mathrm{eV}$
& \cite{Reiner2020Spectroscopic, Schuwalow2019BandBending, Liu2021InAsEuSAl} \\

\hline

EuS
& width
& $4~\mathrm{nm}$
& -- \\

& $m_{\mathrm{eff}}$
& $0.3m_0$
& \cite{Xavier1967Europium} \\

& $E_{\mathrm{F}}$
& $0.7~\mathrm{eV}$
& \cite{Alphenaar2009,Liu2020InAsEuSInterfaces} \\

& $h_{\mathrm{ex}}$
& $0.1~\mathrm{eV}$
& \cite{Mauger1986EuropiumChalcogenides,Alphenaar2009} \\

& $\Delta$
& $0$
& -- \\

& $\alpha$
& $0$
& -- \\

& $\epsilon_{\mathrm{EuS}}$
& $10\epsilon_0$
& \cite{Axe1969Europium} \\

\hline


Dielectrics
& $\mathrm{SiO_2}$ width
& $20~\mathrm{nm}$
& -- \\

& $\mathrm{HfO_2}$ width
& $12~\mathrm{nm}$
& -- \\

\hline

& $\epsilon_{\mathrm{SiO_2}}$
& $3.9\epsilon_0$
& \cite{Robertson2004HighK} \\

& $\epsilon_{\mathrm{HfO_2}}$
& $25\epsilon_0$
& \cite{Levinshtein1996Handbook} \\

& $\epsilon_{\mathrm{vacuum}}$
& $\epsilon_0$
& -- \\
\end{tabular}
\end{ruledtabular}

\end{table}

The simulated geometry is scaled down relative to the experimental device to reduce computational cost, including smaller nanowire, superconducting-shell, and substrate dimensions.
The calculations are therefore intended to capture qualitative trends rather than reproduce the experimental gate-voltage scales quantitatively.
In particular, geometric rescaling modifies the voltages at which transverse subbands become occupied.
Sample-dependent doping, charge disorder, and electrostatic screening further limit a direct correspondence between simulated and experimental gate voltages.

\section{Anomalous Phase Calculations}

The equilibrium energy--phase relation is evaluated using the embedded Matsubara formulation described in Appendix C of the main text.
At each imaginary frequency, the semi-infinite superconducting leads are represented by their exact surface self-energies, and the finite normal segment is eliminated recursively to obtain the phase-dependent boundary matrix $M(i\omega,\varphi)$.
The Matsubara integral for the free-energy difference, $F(\varphi)-F(0)$, is evaluated using Gauss--Legendre quadrature, mapping a finite integration interval onto $\omega\in[0,\infty)$.

For $N$ retained transverse modes, the boundary determinant is a finite Laurent polynomial in $e^{i\varphi}$,
\begin{equation*}
    \det M(i\omega,\varphi)=\sum_{m=-2N}^{2N}c_m(i\omega)e^{im\varphi}.
\end{equation*}
Consequently, $4N+1$ equally spaced phase samples are sufficient to reconstruct the complex determinant exactly at each Matsubara frequency, before taking the logarithm in the free-energy expression.

The calculations retain two transverse hard-wall modes in both the normal segment and superconducting leads, with equal widths $W_{\rm N}=W_{\rm S}=W$.
The semiconductor effective mass is taken as $m^\ast=0.023\,m_0$.
We consider $W=20$, $40$, $60$, $80$, and $100$~nm, with $\Delta=0.2$~meV, $\mu_{\rm S}=10$~meV, $L_{\rm N}=100$~nm, and equal Zeeman-energy vectors $\mathbf B_{\rm N}=\mathbf B_{\rm S}=(0.1,0.1,0)$~meV.
For each width, the normal-region chemical potential is varied from $-0.35$~meV to $0.85\,\epsilon_2(W)$, where $\epsilon_2(W)$ is the second transverse-subband threshold.
The normal segment therefore remains below this threshold, while the retained second mode accounts for intersubband mixing.
In the superconducting leads, the second transverse subband lies below $\mu_{\rm S}$ for $W=80$ and $100$~nm, whereas the third remains above $\mu_{\rm S}=10$~meV for all considered widths.
The dependence of $\vert\varphi_0\vert$ on Rashba coupling and nanowire width in the main-text Fig.~7(c) is evaluated at fixed $\mu_{\rm N}=2$~meV, within the low-density regime common to all five widths.
Unless stated otherwise, energies are expressed in meV, lengths in nm, and the Rashba coefficient in meV\,nm.

\bibliography{Ref}